\documentclass[
    a4paper,
    notitlepage,
    prb,
    twocolumn,
    superscriptaddress,
    floatfix
]{revtex4-1}
\usepackage{amsmath,amssymb}
\usepackage{lipsum}
\usepackage{bm}
\usepackage{graphicx}
\usepackage{graphics}
\usepackage{amsmath}
\usepackage{array}
\usepackage[caption=false]{subfig}
\usepackage{xcolor}
\usepackage{subfig}
\usepackage{hyperref}
\usepackage[capitalise]{cleveref}
\usepackage{textcomp}

\usepackage{epstopdf}
\usepackage{cases}
\usepackage{amssymb}
\usepackage{multirow}
\usepackage{siunitx}
\usepackage{cases}
\usepackage{tabularx}
\usepackage[top=1.1in, bottom=1.1in, left=1.0in, right=1.0in]{geometry}
\usepackage[utf8]{inputenc}
\usepackage{tikz}
\usepackage{comment}
\usepackage{lineno,hyperref}
\modulolinenumbers[5]
\usepackage{booktabs}
\usepackage{diagbox}

\newcommand{\leib}[1]{{\leavevmode\color{black}#1}}

\begin{document}

\title{On the emergence of heat waves in the transient thermal grating geometry}
\author{Chuang Zhang}
\email{zhangc33@sustech.edu.cn}
\affiliation{Department of Mechanics and Aerospace Engineering, Southern University of Science and Technology, Shenzhen 518055, China}
\author{Samuel Huberman}
\email{samuel.huberman@mcgill.ca}
\affiliation{Department of Chemical Engineering, McGill University, 845 Sherbrooke St W, Montreal, Quebec, Canada}
\author{Lei Wu}
\email{Corresponding author: wul@sustech.edu.cn}
\affiliation{Department of Mechanics and Aerospace Engineering, Southern University of Science and Technology, Shenzhen 518055, China}
\date{\today}


\begin{abstract}

{\color{black}{The propagation of heat in the transient thermal grating geometry is studied based on phonon Boltzmann transport equation (BTE) in different phonon transport regimes. Our analytical and numerical results show that the phonon dispersion relation and temperature play a significant role in the emergence of heat wave. For the frequency-independent BTE, the heat wave appears as long as the phonon resistive scattering is not sufficient, while for the frequency-dependent BTE, the heat wave could disappear in the ballistic regime, depending on the grating period and temperature.
We predict that the heat wave could appear in the suspended graphene and silicon in extremely low temperature but disappear at room temperature.}}

\end{abstract}


\maketitle

\section{Introduction}

\begin{figure}[htb]
\centering
\includegraphics[scale=0.35,clip=true]{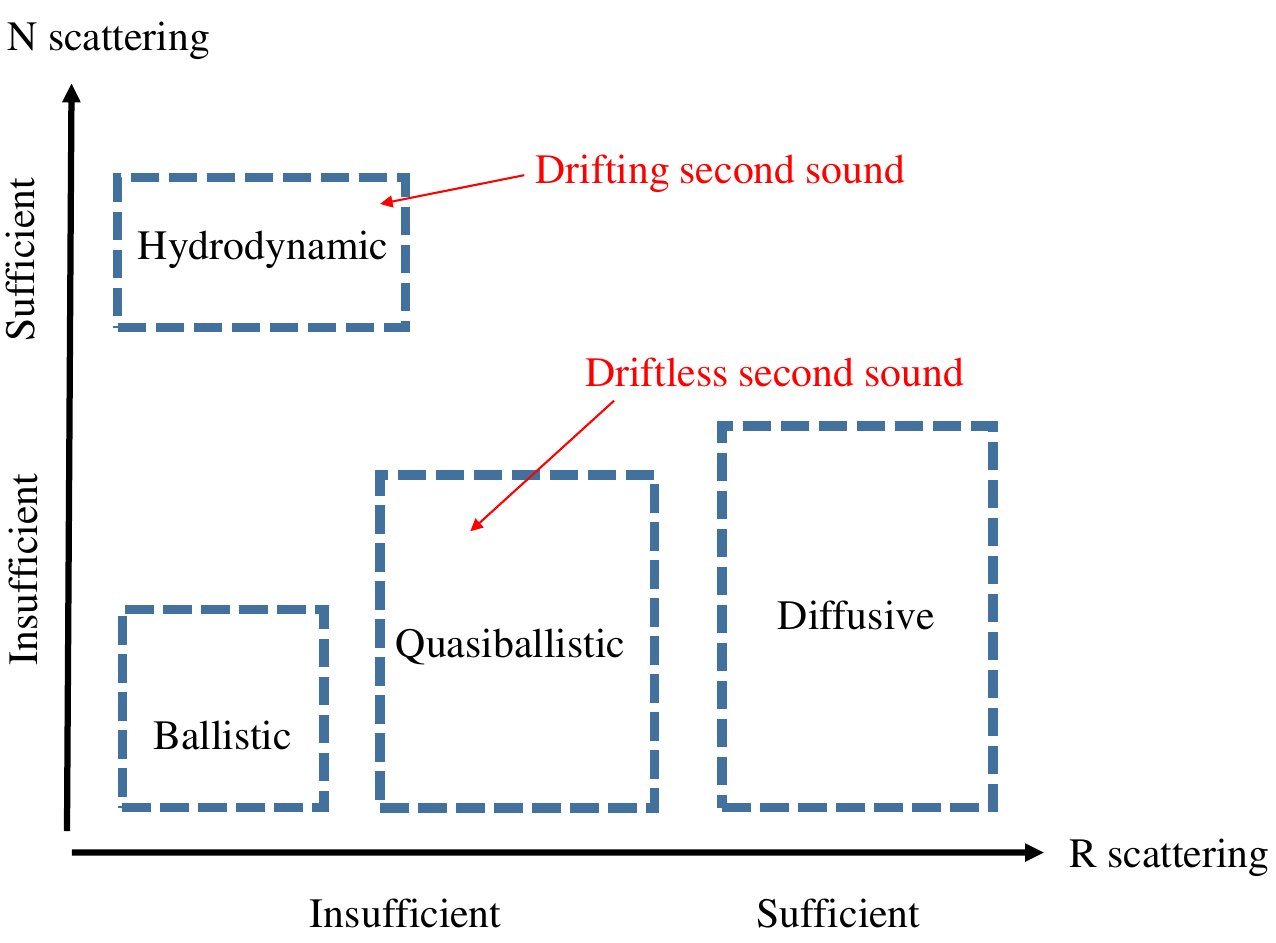}
\caption{A schematic of phonon transport regimes. Note that many terminologies have previously been used to describe the crossover regime among the phonon hydrodynamic, ballistic and diffusive limits~\cite{chen_non-fourier_2021,zhang_transient_2021,leesangyeopch1,PhysRev.148.766,PhysRev_GK,cepellotti_phonon_2015}, such as ballistic-diffusive, quasiballistic and Ziman regime.  }
\label{Quasiballistic}
\end{figure}

The heat wave in solid materials, i.e., the wave like propagation of heat at a finite speed, breaks Fourier's law of thermal conduction~\cite{RevModPhysJoseph89}.
It has been studied over half a century~\cite{ozisik_wave_1994,tzou2014macro,cattaneo1948sulla}, and {\color{black}{three}} main physical mechanisms have been identified~\cite{chen_non-fourier_2021,ENZ1968114,PhysRevB_SECOND_SOUND,leesangyeopch1}, {\color{black}{as shown in~\cref{Quasiballistic}.}}
The first is known as drifting second sound~\cite{PhysRevB_SECOND_SOUND,leesangyeopch1}, which requires both sufficient momentum-conserving phonon normal (N) scattering and insufficient momentum-destroying resistive (R) scattering~\cite{PhysRev.131.2013,Dreyer1993,PhysRev.148.766,PhysRev_GK,PhysRevB.99.085202,gurevichShklov1967seonds}. In this case, hydrodynamic phonon transport dominates heat conduction and the thermal behaviors are similar to fluid dynamics.
The second is known as driftless second sound~\cite{PhysRevB_SECOND_SOUND,ENZ1968114,beardo_observation_2021}, as it does not need sufficient N-scattering~\cite{beardo_observation_2021,PhysRevB_SECOND_SOUND};
however, it requires that all phonons have similar relaxation time and there are external time-dependent heat sources, whose heating period is comparable to the phonon relaxation time~\cite{PhysRevB.101.075303}.
Thus, phonon scattering is not sufficient in one heating period and {\color{black}{quasiballistic phonon transport}} dominates the transient heat conduction.
{\color{black}{The third is ballistic heat propagation, which appears when the system characteristic length is much smaller than the phonon mean free path or the system characteristic time is much shorter than the phonon relaxation time~\cite{GuoZl16DUGKS,kovacs2018,PhysRevB.99.085202,collins_non-diffusive_2013,zhang_transient_2021}.
In the ballistic regime, there is no phonon intrinsic scattering.}}

Many of the past studies of heat waves focused on the drifting second sound, which was first observed in solid helium~\cite{PhysRevLett.16.789}.
Initially, drifting second sound was measured in a few materials, such as  NaF~\cite{PhysRevLett_secondNaF,PhysRevLett_ssNaf},  Bi~\cite{PhysRevLett.28.1461} and SrTiO$_3$~\cite{PhysRevLett.99.265502}, at extremely low temperatures (near $10$ K).
However, with the development of low-dimensional materials and modern experimental techniques, interest in drifting second sound has experienced a resurgence~\cite{chen_non-fourier_2021,lee2017,luo2019,leesangyeopch1,RevModPhys.90.041002}.
In 2015, Lee \textit{et al.} found that the intensity of N-scattering in suspended graphene can be significantly greater than that of the R-scattering ~\cite{lee_hydrodynamic_2015} and phonon properties obtained from first principles calculations as input to the phonon Boltzmann transport equation (BTE), they predicted the occurrence of drifting second sound at $100$~K. Four years later, drifting second sound was observed in graphite by transient thermal grating experiments at temperatures above $100$~K ~\cite{huberman_observation_2019}.

In comparison to drifting second sound, there have only been a handful of studies that focused on driftless second sound~\cite{ENZ1968114,PhysRevB_SECOND_SOUND,cepellotti_phonon_2015,beardo_observation_2021} in the quasiballistic regime.
In 2015, Cepellotti \textit{et al.} theoretically studied the driftless second sound in some two-dimensional materials including graphene and boron nitride~\cite{cepellotti_phonon_2015}. The frequency-domain thermoreflectance experiment has been proposed as a platform to observe driftless second sound if the heating frequency is sufficiently high compared to the inverse of the phonon relaxation time~\cite{PhysRevB.101.075303}.
Very recently, driftless second sound has been observed in Germanium between $7$ K and room temperature~\cite{beardo_observation_2021} in a frequency-domain thermoreflectance experiment. The phase lag between the heating frequency and the temperature response cannot be modeled by a modified Fourier's law, but requires a hyperbolic heat equation ~\cite{beardo_observation_2021}.

{\color{black}{In the past, the relative intensity of phonon normal, resistive and boundary scattering has been a helpful guide in distinguishing (quasi)ballistic phonon transport and phonon hydrodynamics~\cite{chen_non-fourier_2021,ENZ1968114,beardo_observation_2021,PhysRevB_SECOND_SOUND}.
However, any of these phenomena can manifest as heat waves at the macroscopic level.}}
For example, when a simple linear phonon dispersion is considered, drifting and driftless second sound have the same propagation speed~\cite{PhysRevB_SECOND_SOUND}.
Hence an open question naturally arises, that is, when a heat wave is experimentally observed, how can we ascertain which corresponding regime of transport is taking place ~\cite{zhang_transient_2021,PhysRevB_SECOND_SOUND}?
{\color{black}{Before clearly distinguishing whether an observed heat wave is due to quasiballistic or hydrodynamic processes, it is necessary to study heat waves in a number of thermal systems, materials and phonon transport regimes.}}

So far, only a few studies focused on the macroscopic signatures of heat wave in different phonon transport regimes~\cite{kovacs2018,huberman_observation_2019,zhang_transient_2021}.
In 2019, Huberman \textit{et al.} studied the heat wave in graphite with different grating periods and temperatures through transient thermal grating (TTG) experiments and linearized phonon BTE simulations~\cite{huberman_observation_2019}.
For the temperatures and lengths which were probed, they found that the signal of negative dimensionless temperature can only be observed in the hydrodynamic regime.
Recently, through the quasi-two and three dimensional numerical simulations with various N/R scattering, Zhang \textit{et al.} found the transient temperature can be lower than the lowest value of initial temperature in the hydrodynamic regime, within a certain range of time and space~\cite{zhang_transient_2021}.
This novel phenomenon of heat wave appears only when the N-scattering dominates the heat conduction, but disappears in quasi-one dimensional simulations. This effect was recently measured in graphite in the hydrodynamic regime by picosecond laser irradiation~\cite{PhysRevLett.127.085901}.

{\color{black}{In order to obtain a clearer understanding of the microscopic origins of the macroscopic signatures of heat waves in various phonon transport regimes, we study the TTG in two- and three-dimensional materials based on solving the phonon BTE numerically and analytically, in the effort to provide theoretical guidance for future experiments~\cite{huberman_observation_2019,beardo_observation_2021,ding_observation_2022}.}}


\section{The phonon BTE}
\label{sec:bte}

In order to describe transient heat propagation in different transport regimes, the phonon BTE under the Callaway approximation is used~\cite{ChenG05Oxford,kaviany_2008,PhysRev_callaway,wangmr17callaway,luo2019},
\begin{align}\label{eq:BTE}
\frac{\partial e}{\partial t }+ v_g \bm{s} \cdot \nabla_{\bm{x}} e  &= \frac{e^{eq}_{R} -e}{\tau_{R}} + \frac{e^{eq}_{N}-e}{\tau_{N}}   ,
\end{align}
where $e=e(\bm{x},\omega,\bm{s},t,p)$ is the phonon distribution function, which depends on spatial position $\bm{x}$, unit directional vector $\bm{s}$, time $t$, phonon frequency $\omega$, and polarization $p$.
$v_g$ is the group velocity, which can be calculated by the phonon dispersion (see Appendix~\ref{sec:dispersionsilicon}).
$e^{eq}_R$ and $e^{eq}_N$ are the equilibrium distribution functions of the R-scattering and N-scattering, respectively~\cite{ChenG05Oxford,kaviany_2008}, while $\tau_R$ and $\tau_N$ are the corresponding relaxation times to which the distribution function $e$ relaxes.

We assume the temperature $T$ slightly deviates from the reference temperature $T_0$, i.e., $|T -T_0| \ll  T_0$, so that the equilibrium distribution function can be linearized as follows:
\begin{align}
e^{eq}_{R}(T) &\approx C \left( T-T_0 \right) / B,  \label{eq:feqR} \\
e^{eq}_{N}(T,\bm{u}) &\approx  C \left( T-T_0 \right) / B +CT \frac{\bm{K} \cdot \bm{u} }{ B \omega}, \label{eq:feqN}
\end{align}
where $C=C(\omega,p,T_0)$ is the mode specific heat at $T_0$, $\bm{u}$ is the drift velocity, and $\bm{K}$ is the isotropic wave vector.
{\color{black}{The specific heat, group velocity and relaxation time all depend on the reference temperature $T_0$.}}
The local temperature $T$ and heat flux $\bm{q}$ can be calculated as the moments of distribution function:
\begin{align}
T  &=T_0+ \frac{ \sum_{p} \int \int e d\Omega d\omega } { \sum_{p} \int C d\omega  },   \label{eq:T}  \\
\bm{q} &=  \sum_{p} \int \int \bm{v} e d\Omega d\omega ,
\label{eq:heatflux}
\end{align}
where the integral is carried out  in the whole solid angle space $d\Omega$  and frequency space $d\omega$. For two- and three-dimensional materials, we have $B=2\pi$ and $4\pi$, respectively.

{\color{black}{
R scattering satisfies the energy conservation while N scattering satisfies both the energy and momentum conservation. Therefore, we have the following equations:
\begin{align}
0  &=  \sum_{p} \int \int  \frac{e^{eq}_{R}(T_R) -e}{\tau_{R}(T_0)}   d\Omega d\omega , \label{eq:Rconsertvation}   \\
0  &=  \sum_{p} \int \int    \frac{e^{eq}_{N} (T_N, \bm{u})-e}{\tau_{N}(T_0)}   d\Omega  d\omega , \label{eq:Nconsertvation}  \\
0 &=  \sum_{p} \int \int  \frac{\bm{K} }{\omega } \frac{e^{eq}_{N}(T_N,\bm{u})-e}{\tau_{N}(T_0)}   d\Omega d\omega,
\label{eq:NMconsertvation}
\end{align}
where $T_R$ and $T_N$ are local pseudotemperatures, which are introduced to ensure the conservation principles of the phonon scattering.
Combining the above three equations, the macroscopic variables $T_R$, $T_N$ and $\bm{u}$ can be obtained,
\begin{align}
T_{R} & =T_{0}+ \left( \sum_{p}\int  \frac{\int e d\Omega }{\tau_R } d{\omega}   \right) \times \left(  \sum_{p}\int \frac{ C}{\tau_R }  d{\omega} \right)^{-1} , \\
T_{N} & =T_{0}+ \left( \sum_{p}\int  \frac{\int e d\Omega }{\tau_N } d{\omega}   \right) \times \left(  \sum_{p}\int \frac{ C}{\tau_N }  d{\omega} \right)^{-1} , \\
\bm{u}  &=  \frac{A}{T_{N}  \sum_{p} \int \frac{ |\bm{K}|^2  }{ \omega^2  } \frac{ C }{\tau_N } d{\omega}  } \sum_{p} \int \int  \frac{ \bm{K} }{\omega} \frac{ e }{ \tau_N}   d\Omega d\omega,
\end{align}
where $A=2$ for two-dimensional materials and $A=3$ for three-dimensional materials.
}}

{\color{black}{
In this study, the frequency-independent and frequency-dependent phonon BTE are used.
For the former, all phonon polarizations and frequencies are the same, and there is only single group velocity, specific heat and relaxation time~\cite{zhang_transient_2021,shang_heat_2020,collins_non-diffusive_2013}.
In this case, we can derive $ A \bm{q} = C T \bm{u} $~\cite{WangMr15application,zhang_transient_2021}, $T = T_R =T_N$, and the momentum conservation is equivalent to the heat flux conservation.
For a more realistic model of materials, the frequency-dependent phonon BTE is solved with frequency-dependent group velocity, specific heat and relaxation time~\cite{wangmr17callaway,luo2019}. In this case,
$T \neq T_R \neq T_N$ and the momentum conservation is not equivalent to the heat flux conservation~\cite{leesangyeopch1,cepellotti_phonon_2015}.
}}

The phonon BTE~\eqref{eq:BTE} can be rewritten in dimensionless form:
\begin{align}\label{eq:dimensionlessBTE}
\frac{\partial e}{\partial t }+ \bm{s} \cdot \nabla_{\bm{x}} e  &= \frac{e^{eq}_{R} -e}{\text{Kn}_{R} }  + \frac{e^{eq}_{N} -e}{\text{Kn}_{N}},
\end{align}
where the distribution function is normalized by $e_{\text{ref}}  ={ C  \Delta T  }/{ B }$ with $\Delta T$ being the temperature difference in the domain, the spatial coordinates normalized by $L$, and time normalized by $t_{ref}=L/v_g $.
The two dimensionless Knudsen numbers are
\begin{equation} \label{eq:dimensionlessparameters}
\text{Kn}_{R}^{-1} = \frac{L}{v_g \tau_R },
\quad
\text{Kn}_{N}^{-1} = \frac{ L  }{v_g \tau_N},
\end{equation}
which index the strength of R- and N-scattering, respectively. The limits of
$\text{Kn}_R^{-1} \ll 1 \ll  \text{Kn}_N^{-1} $ corresponds to the hydrodynamic regime. The limits of $\text{Kn}_R^{-1} \ll  1$ and $\text{Kn}_N^{-1}  \ll  1$ correspond to the ballistic regime.
{\color{black}{When N scattering is not sufficient, and the heat conduction is between the ballistic and diffusive limits, we classify it as the quasiballistic regime~\cite{leesangyeopch1}, as shown in~\cref{Quasiballistic}.        }}

\section{Results and discussions}
\label{sec:resultsBTE}

Our heating geometry is equivalent to the quasi-one dimensional transient thermal grating experimental geometry~\cite{collins_non-diffusive_2013}.
At the initial moment $t=0$, there is a spatially cosine temperature distribution in the quasi-one dimensional domain:
\begin{align}
T(x,t=0)=T_0+ \Delta T \cos( w x ),
\label{eq:initialBC}
\end{align}
where  $w= 2 \pi/ L$ is the spatial wave vector, $L$ is the spatial grating period.
The periodic boundary conditions are used and we focus on the transient variations of temperature at position $x=0$.
In the diffusive regime, the heat conduction follows the Fourier law, and the temperature decays exponentially with time~\cite{collins_non-diffusive_2013}:
\begin{align}\label{eq:diffusive}
T= T_0 + \Delta T \cos( w x ) \exp{ \left(-\alpha  w^2 t  \right)},
\end{align}
where $\alpha=\kappa /C$ is the thermal diffusivity with $\kappa$ being the thermal conductivity.
The temperature profile for heat transport in two- and three-dimensional materials is obtained both via numerical methods detailed in Appendix~\ref{sec:solver} and via analytical solutions presented below.


\textbf{Frequency Independent Ballistic regime}---$\text{Kn}_R^{-1} \ll 1$ and $\text{Kn}_N^{-1} \ll 1$. Phonons transport occurs without intrinsic scattering, so
$e(\bm{x}, \bm{s}, t+\delta t )= e(\bm{x}- v_g \bm{s} \delta t, \bm{s}, t  )$,
where $\delta t$ is an arbitrary time interval.
Then, according to Eq.~\eqref{eq:T}  we have
\begin{equation}
\begin{aligned}
T(\bm{x},t)=T_0 + \frac{1}{C}  \int e(\bm{x}-v_g \bm{s} t,\bm{s}, 0) d\Omega.
\end{aligned}
\end{equation}
At the initial moment, all phonons follow the  equilibrium distribution  $e^{eq}_R (T)$. Therefore, the dimensionless temperature
\begin{equation}\label{dimensionless_temp}
T^* =\frac{T-T_0}{ \Delta T}
\end{equation}
in two-dimensional materials is
\begin{equation} \label{eq:TTG2Dballistic}
\begin{aligned}[b]
T^*&= \frac{1}{2 \pi}  \int_0^{2 \pi} \cos \left( 2 \pi v_g t/L \cos \theta \right)  d \theta  \\
&=    J_0 (2 \pi v_g  t /L ),
\end{aligned}
\end{equation}
where $J_0$ is the zeroth order Bessel function. In the three-dimensional materials, the analytical solution in the ballistic regime is
\begin{equation}\label{eq:TTG3Dballistic}
\begin{aligned}
 T^* = \frac{ \sin{\left(  2 \pi v_g t/L \right)}  }{2 \pi v_g t/L  }.
\end{aligned}
\end{equation}

\textbf{Frequency Independent Phonon Hydrodynamic regime}--$\text{Kn}_R^{-1} \ll 1 \ll  \text{Kn}_N^{-1} $. The distribution function can be approximated based on Chapman-Enskog expansion~\cite{chapman1995book} as $e =  e_N^{eq} - \tau_N  \left( \frac{ \partial e_N^{eq}}{\partial t} + v_g \bm{s} \cdot \nabla e_N^{eq}    \right) + O(\tau_N^2)$. Taking the zero- and first- order moments of phonon BTE~\eqref{eq:BTE}, we have
\begin{equation}
\begin{aligned}[b]
C \frac{ \partial T}{ \partial t } + \nabla \cdot \bm{q} &=0,  \\
\frac{ \partial \bm{q} }{ \partial t } + \nabla \cdot \bm{Q} &=0,
\end{aligned}
\end{equation}
where
\begin{equation*}
\begin{aligned}[b]
\bm{Q} &=\int \bm{v}\bm{v} e d\Omega\\
&\approx \int \bm{v}\bm{v}  \left[ e_N^{eq} - \tau_N  \left( \frac{ \partial e_N^{eq}}{\partial t} + v_g \bm{s} \cdot \nabla e_N^{eq}    \right)    \right] d\Omega      \\
&= a  CT +  b  C  \frac{ \partial  T}{ \partial t },
\end{aligned}
\end{equation*}
with $a=\frac{v_g^2}{3}$ and $b=\frac{4}{15}  v_g^2 \tau_N $ in  three-dimensional materials,  and  $a=\frac{v_g^2}{2}$ and $b= \frac{v_g^2 \tau_N }{4}$ in two-dimensional materials.
{\color{black}{The drift velocity $ \bm{u} = 3 \bm{q}/(CT )$ is replaced by the heat flux in above derivations~\cite{WangMr15application,zhang_transient_2021,beck1974}, and the spatial divergence of the heat flux is replaced by the temporal derivation of the temperature due to the energy conservation.}}
Note that in the phonon hydrodynamic regime, $b \rightarrow 0$.
Therefore, the temperature evolution is governed  by
\begin{align}
\frac{ \partial^2 T}{ \partial t^2 } - a   \nabla^2 T  -  b   \frac{ \partial (\nabla^2 T) }{ \partial t}  =0,
\label{eq:hyperbolicheatTTG}
\end{align}
and under the initial condition~\eqref{eq:initialBC}, the dimensionless temperature in two-dimensional materials is given by the following equation
\begin{align}
T^*  \approx \exp{\left( - \frac{ \pi ^2  v_g^2  \tau_N  }{ 2L^2 } t
    + i \frac{2 \pi }{L} \left(x  \pm  \frac{v_g}{\sqrt{2}}  t  \right)    \right)  },
\label{eq:analyhydroTTG2D}
\end{align}
where $i$ is the imaginary number.
From Eq.~\eqref{eq:analyhydroTTG2D} it is seen that the undamped heat wave occurs only when $\tau_N =0$, and its propagation speed is $v_g/ \sqrt{2}$. In the three-dimensional materials, the analytical solution in the hydrodynamic regime is
\begin{align}
T^* \approx  \exp{\left( - \frac{8 \pi ^2 v_g^2  \tau_N  t}{15 L^2 } + i \frac{2 \pi }{L} (x  \pm  \frac{v_g}{\sqrt{3}}  t  )    \right)  }.
\label{eq:analyhydroTTG3D}
\end{align}

\subsection{Two-dimensional materials}
\label{sec:TTG1D}

\subsubsection{Frequency-independent system}

\begin{figure}[t]
\centering
\includegraphics[scale=0.35,clip=true]{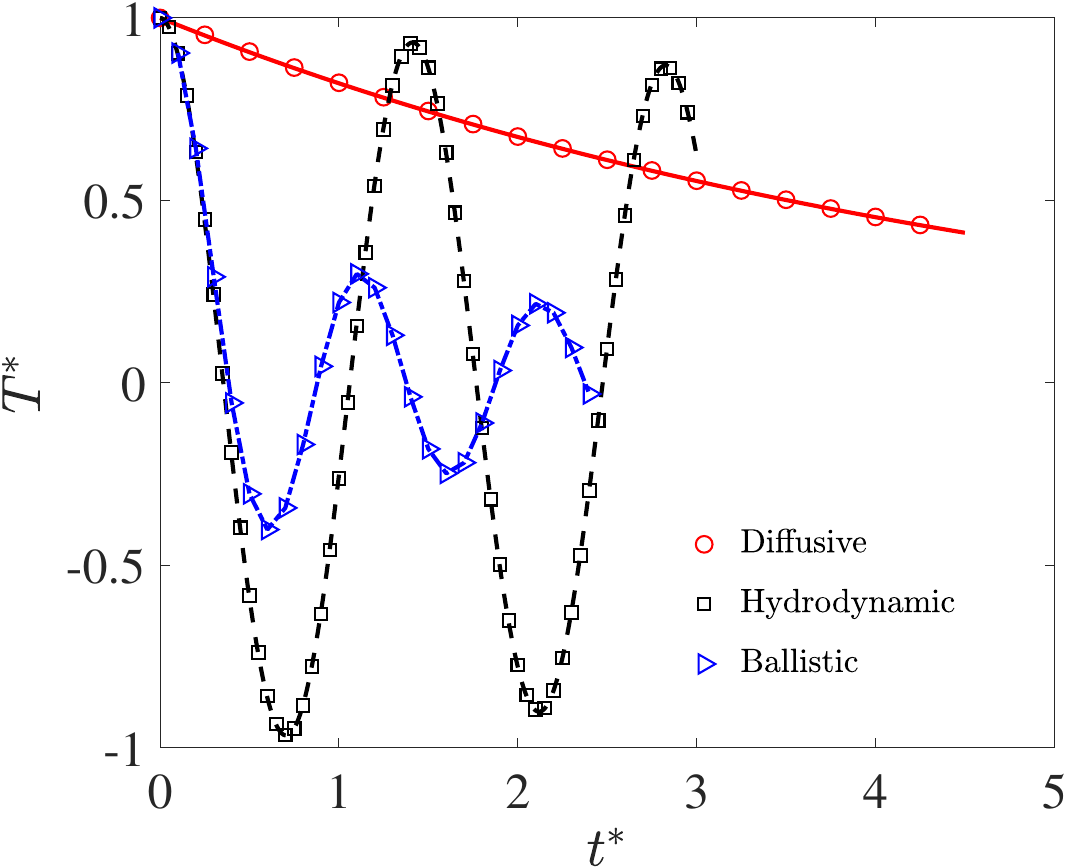}
\caption{The evolution of the dimensionless temperature~\eqref{dimensionless_temp} in the TTG simulations of two-dimensional materials based on frequency-independent phonon BTE, where $t^*=v_g t/L$. \leib{Symbols: numerical solutions in the diffusive ($\text{Kn}_R^{-1}=100$, $\text{Kn}_N^{-1}=0$), ballistic ($\text{Kn}_R^{-1}=0$, $\text{Kn}_N^{-1}=0$) and hydrodynamic ($\text{Kn}_R^{-1}=0$, $\text{Kn}_N^{-1}=100$) regimes  based on the frequency-independent phonon BTE. Lines: analytical solutions in the diffuse, ballistic, and hydrodynamic regimes  described by  Eqs.~\eqref{eq:diffusive} with thermal conductivity $\kappa=Cv_g^2 \tau_R/2$, Eq.~\eqref{eq:TTG2Dballistic}, and Eq.~\eqref{eq:analyhydroTTG2D}, respectively. } }
\label{TTGBD}
\end{figure}
\begin{figure*}[t]
\centering
\subfloat[R-scattering only] { \includegraphics[scale=0.28,clip=true]{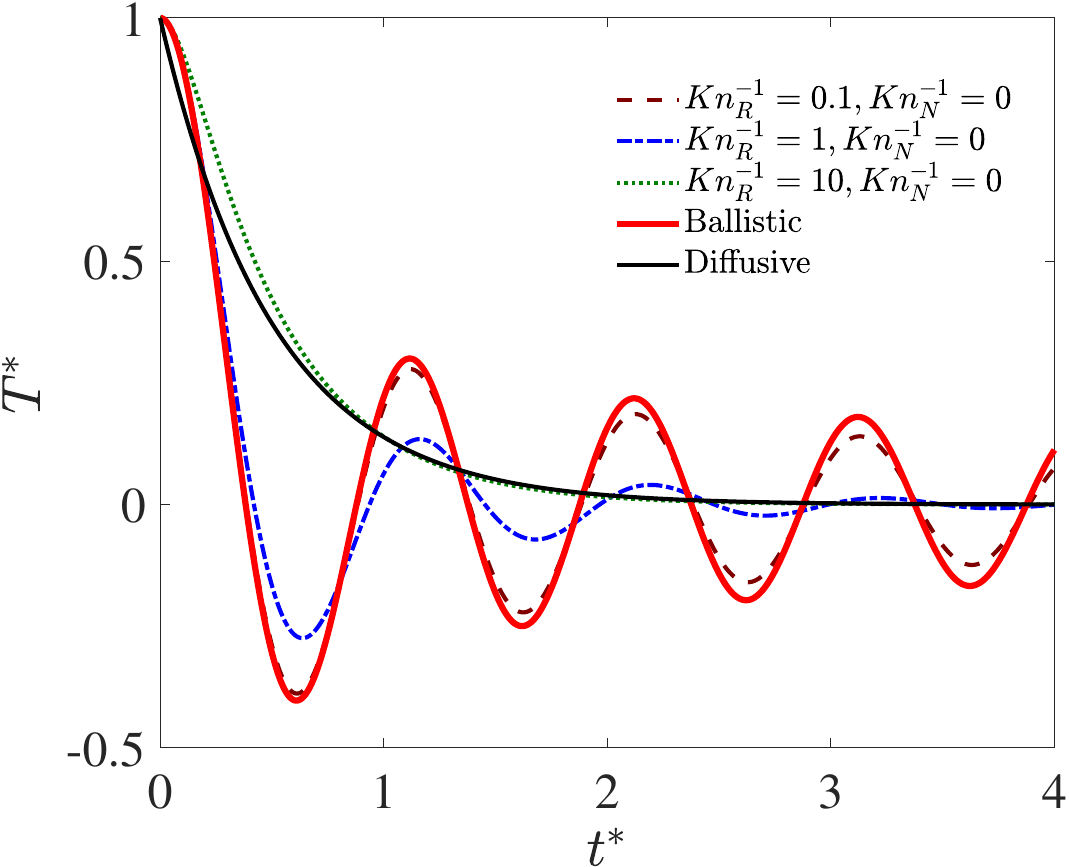} }
\subfloat[N-scattering only] { \includegraphics[scale=0.28,clip=true]{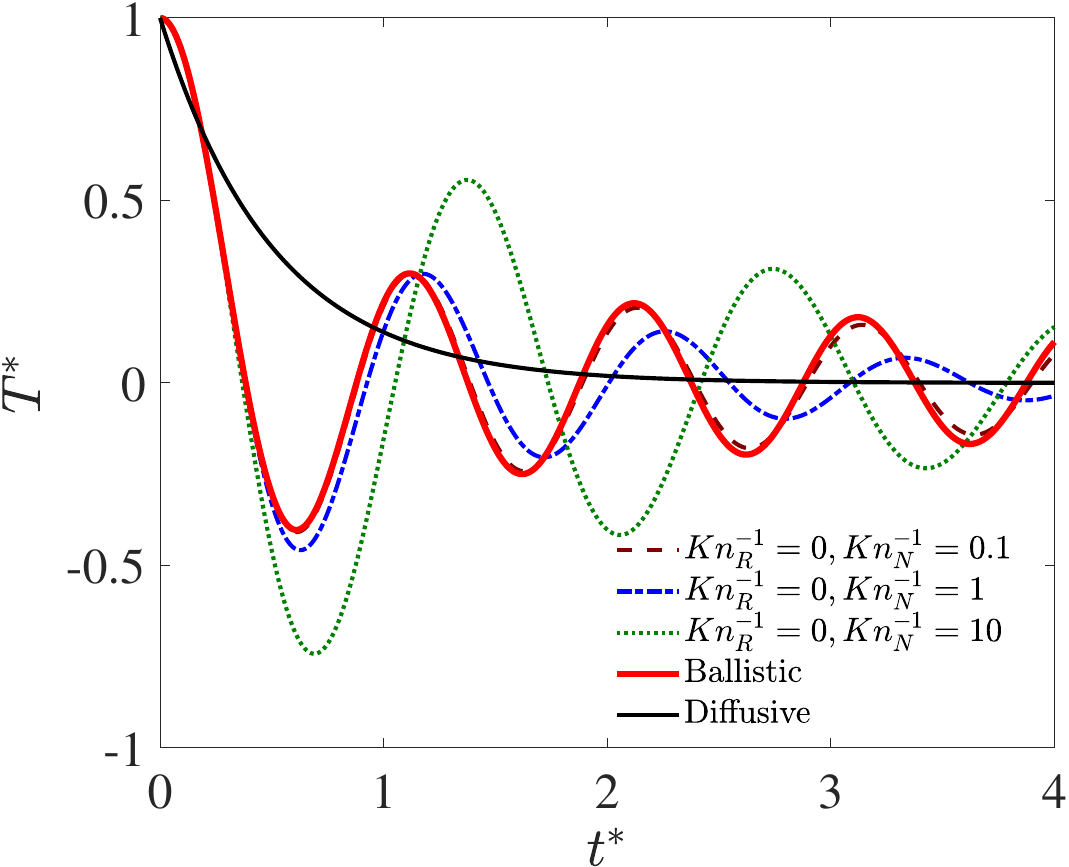} }
\subfloat[Both N/R-scatterings] { \includegraphics[scale=0.28,clip=true]{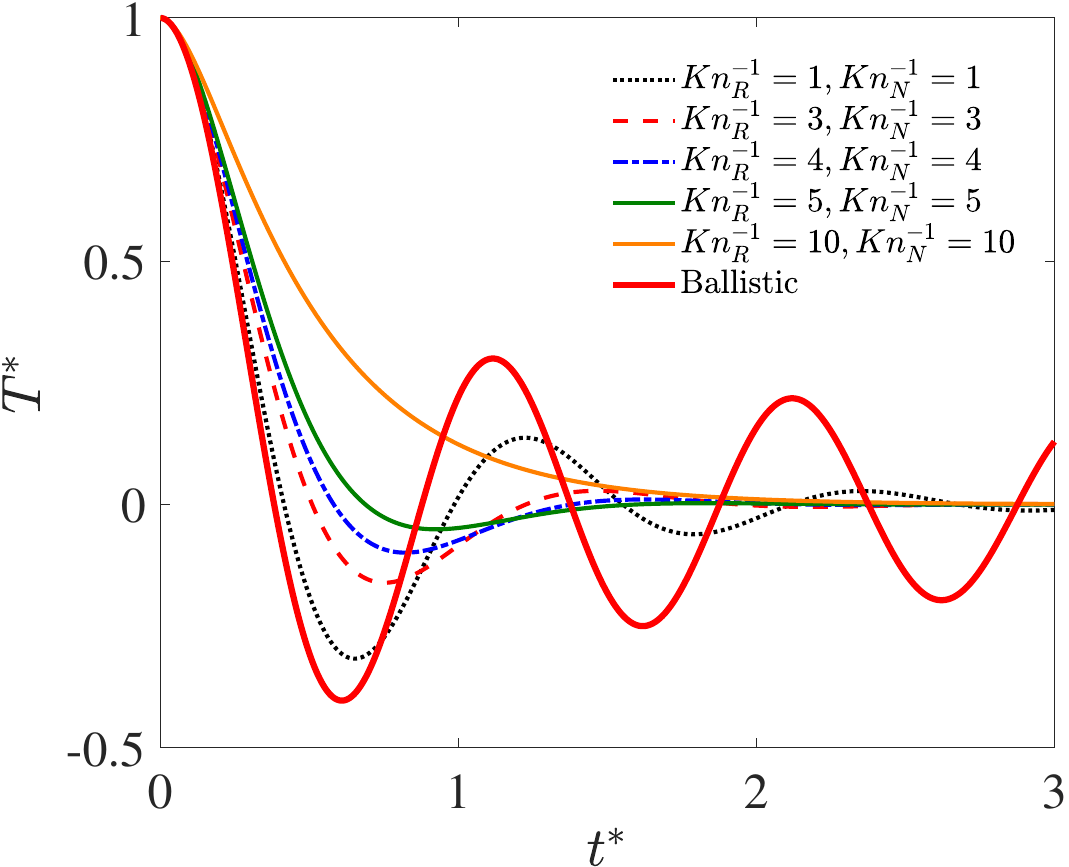} }
\caption{Same as Fig.~\ref{TTGBD}, but with different combinations of $\text{Kn}_R^{-1}$ and $\text{Kn}_N^{-1}$.}
\label{TTGgray}
\end{figure*}

In this subsection, the frequency-independent phonon BTE is used to illustrate the transient heat conduction, since it is useful to pinpoint the essential effects of phonon N- and R-scattering on transient thermal conduction~\cite{shang_heat_2020,collins_non-diffusive_2013}.

Our analytical results in different phonon transport regimes are shown in~\cref{TTGBD}.
In the diffusive regime, the temperature decays exponentially with time as per Eq.~\eqref{eq:diffusive}.
However, this behavior is not observed in both the ballistic and hydrodynamic regimes. Instead, the dimensionless temperature
could become negative, which indicates that the transient temperature at $x=0$ changes from the crest to the trough.
That is, the heat wave appears in both the ballistic and hydrodynamic regimes.

\begin{figure*}
	\centering
	\subfloat[]{\includegraphics[scale=0.28,clip=true]{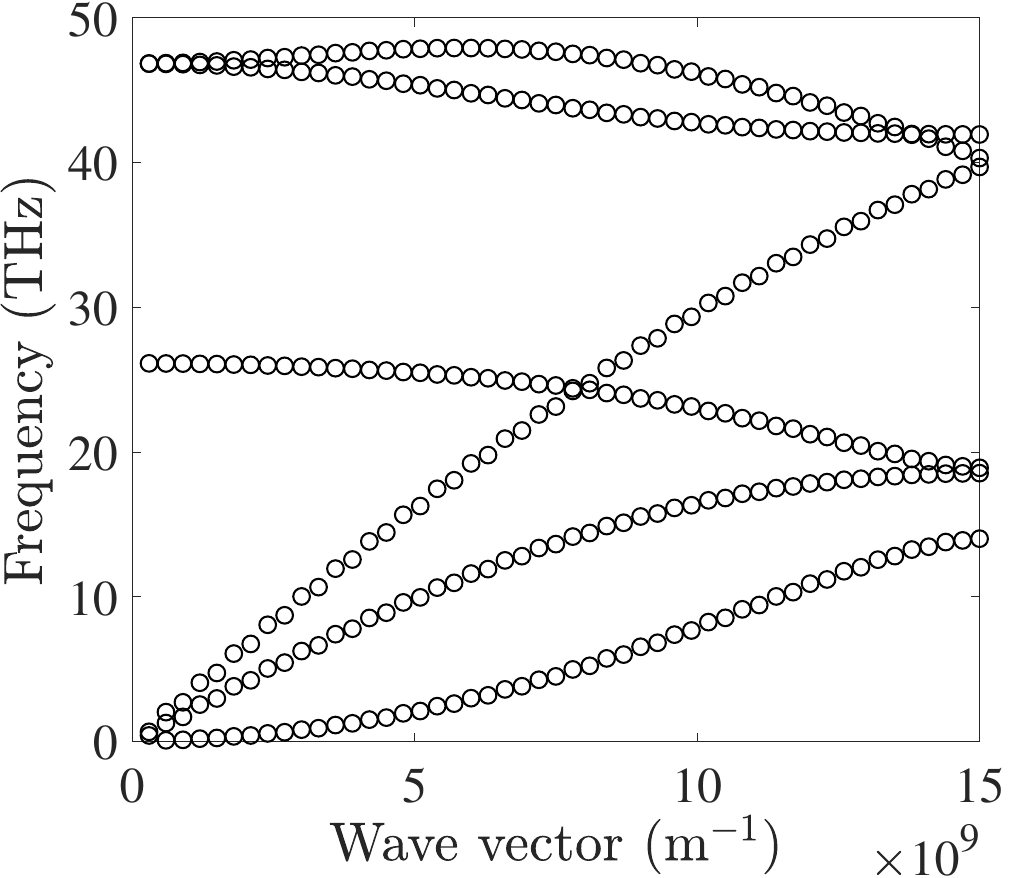} }~~
	\subfloat[]{\includegraphics[scale=0.28,clip=true]{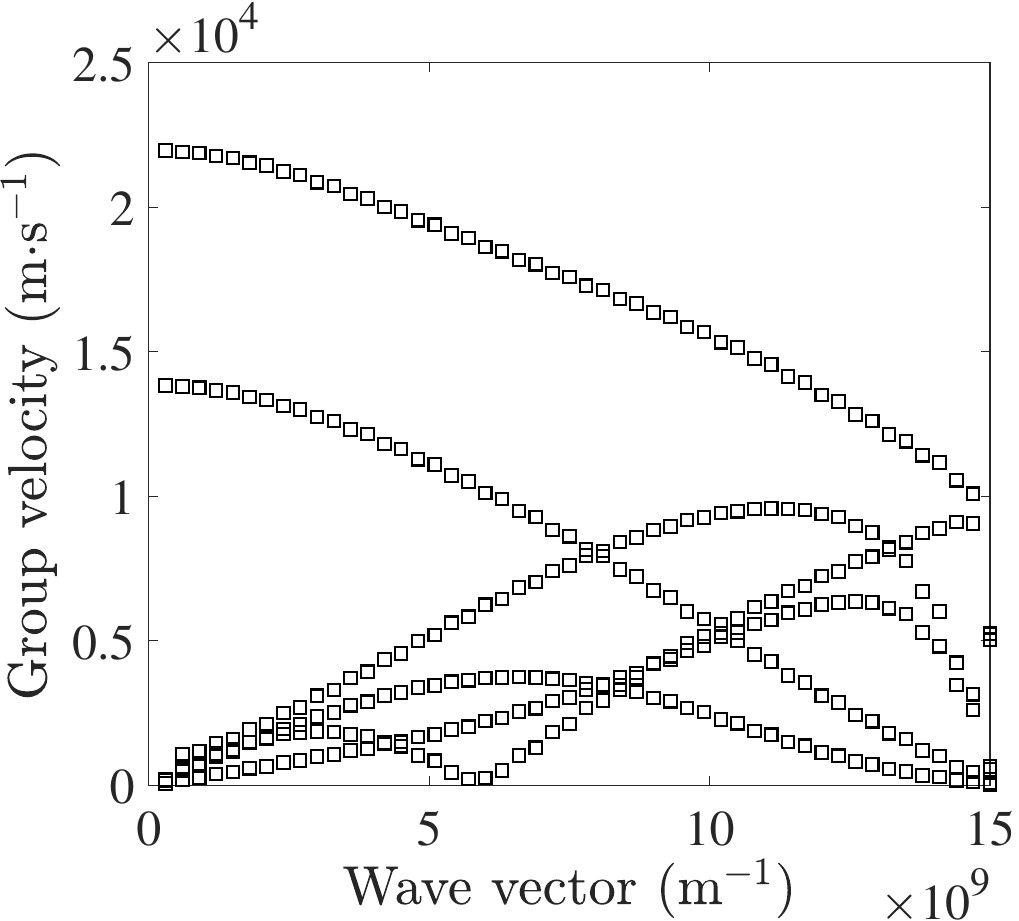}}~~
	\subfloat[]{\includegraphics[scale=0.28,clip=true]{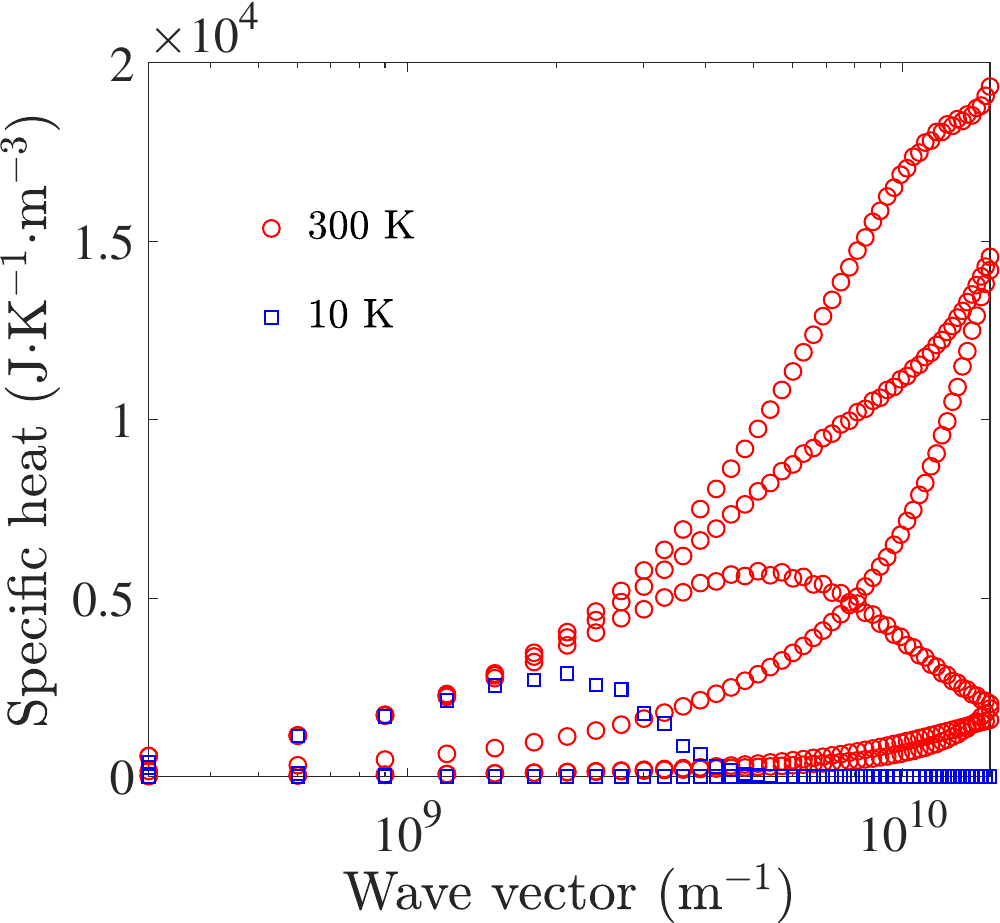}}
	\caption{
		Properties of the single-layer suspended graphene calculated by VASP. (a) Phonon dispersion along $\Gamma$-M direction. (b) Group velocity. (c) Specific heat. 
	}
	\label{garpheneparameters}
\end{figure*}


\textbf{Numerical simulations}---Figure~\ref{TTGBD} demonstrates the accuracy of our numerical method in the diffusive, ballistic, and hydrodynamic regimes. Now, the effects of N- and R-scatterings on the heat wave are investigated in the crossover regime.
For the temperature wave curves, when it reaches the first trough by a period of time $t_1$, we say that the temperature wave has traveled half a wavelength $L/2$~\cite{huberman_observation_2019}.
Thus, the speed of heat propagation is calculated as $v_p = L/(2 t_1)$; one example is shown in~\cref{TTGBD} in the hydrodynamic regime, where one can find that the time difference between two neighboring crests is $\sqrt{2}L/v_g$.

In the absence of N-scattering, when the R-scattering increases so that the phonon transport is from the ballistic to diffusive regime, the heat wave gradually disappears and its propagation speed decreases from $v_g$ to $0$, see Fig.~\ref{TTGgray}(a).
When the R-scattering is absent, the phonon scattering satisfies the momentum conservation, and Fig.~\ref{TTGgray}(b) shows that the heat wave always exists irrespective of the intensity of N-scattering.
That is to say, the heat wave always appears from ballistic to hydrodynamic regime and its propagation speed decreases from $v_g$ to $v_g/ \sqrt{2}$.
With different combinations of N/R scattering, the propagation speed of heat wave could be any value between $0$ and $v_g$, see Fig.~\ref{TTGgray}(c).
In a word, the heat wave could appear in TTG geometry {\color{black}{as long as the R scattering is not sufficient}} based on frequency-independent phonon BTE.

\subsubsection{Graphene}

The transient heat conduction in single-layer suspended graphene (the natural abundance is $1.1\%$~$^{13}$C) is studied based on the frequency-dependent BTE~\eqref{eq:dimensionlessBTE}, where the phonon dispersion and polarization are calculated by the Vienna \textit{Ab initio} Simulation Package (VASP) combined with phonopy, see~\cref{garpheneparameters}.
The thermal conductivity of graphene with infinite size predicted by Callaway’s model~\cite{wangmr17callaway,PhysRev_callaway} is $2674$ W/(m·K) at $300$ K.
Details of these calculations can be found in Ref.~\cite{chuang2021graded}.

\begin{figure}[htb]
\centering
\includegraphics[scale=0.35,clip=true]{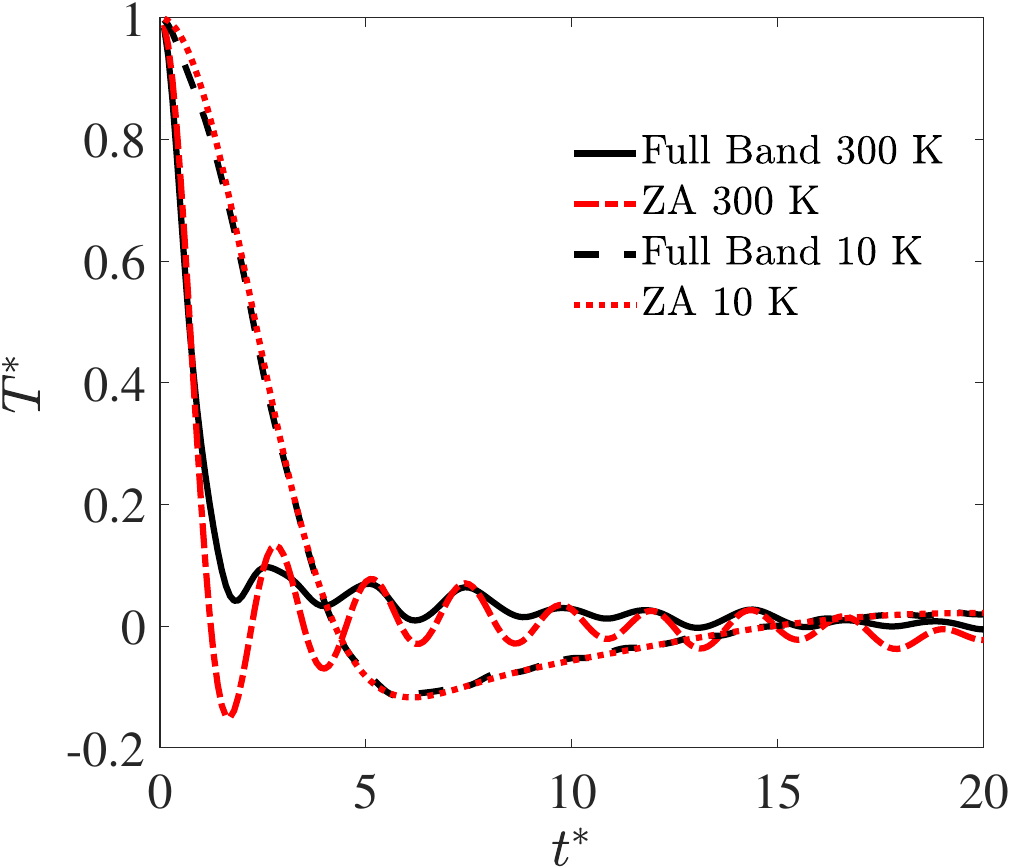}
\caption{The evolution of the dimensionless temperature~\eqref{eq:TTGgraballistic} in graphene in the ballistic regimes.  `Full band' represents that the full six phonon branches in graphene are accounted. `ZA' represents that only phonon  ZA branch is accounted, see \cref{garpheneparameters}(a). The time is $t^*=v_{\text{max}} t/L$ with $v_{\text{max}}=2.2 \times 10^4$ m/s being the maximum group velocity.  }
\label{TTGgarpheneBallistic}
\end{figure}
\begin{figure}[htb]
\centering
\subfloat[$L=10~\mu$m]{\includegraphics[scale=0.35,clip=true]{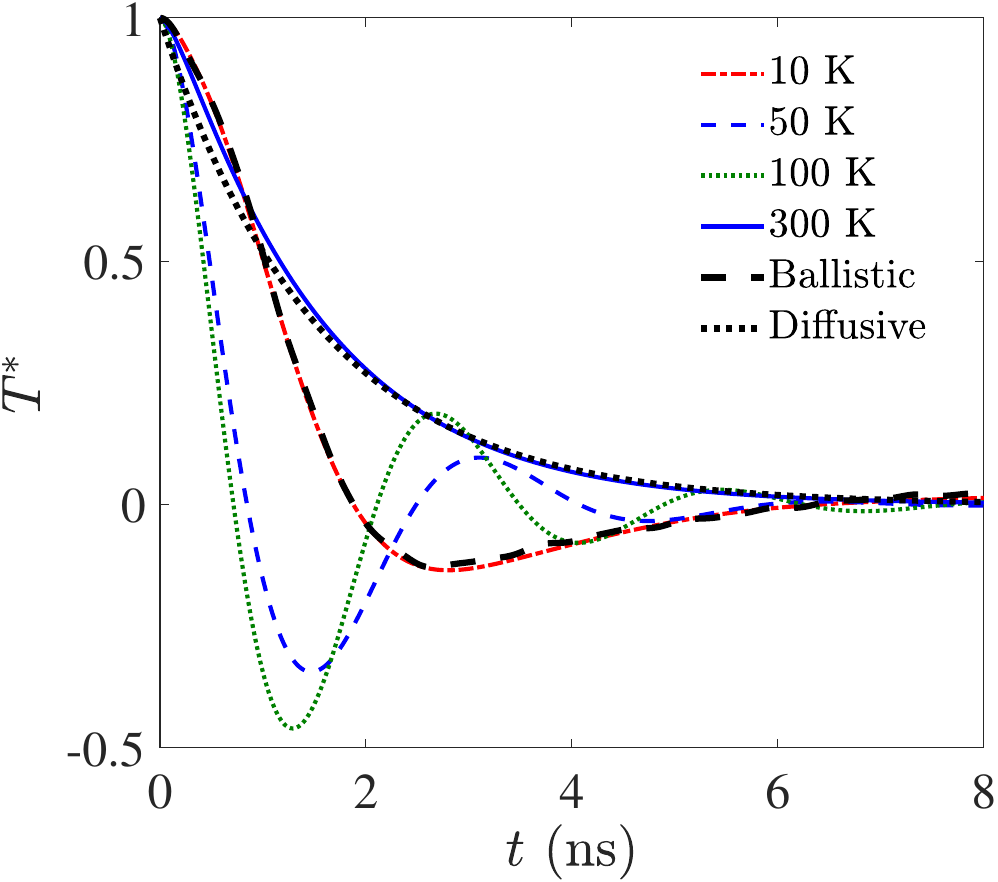}} \\
\subfloat[$L=1~\mu$m]{\includegraphics[scale=0.35,clip=true]{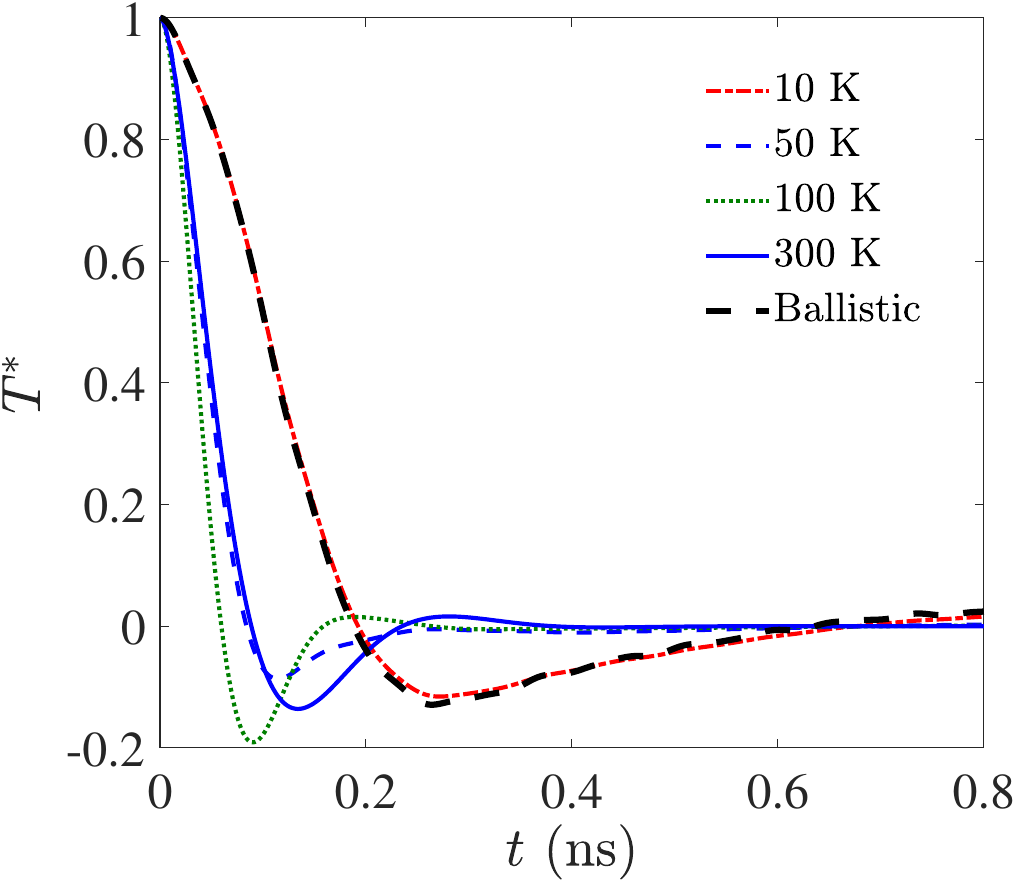}}
\caption{The evolution of the dimensionless temperature~\eqref{dimensionless_temp} in graphene with different reference temperatures $T_0$ and grating period $L$. Ballistic: Eq.~\eqref{eq:TTGgraballistic}.  Diffusive:  Eq.~\eqref{eq:diffusive} with the thermal diffusivity $0.0017$ m$^2$s$^{-1}$. }
\label{TTGgarphene}
\end{figure}
Similar to the procedure used to find the analytical solution~\eqref{eq:TTG2Dballistic} of the frequency-independent BTE, an analytical solution in the ballistic regime can be found:
\begin{align}
T^*  =  \frac{ \sum_p \int  C J_0 (2 \pi v_g  t /L)  d \omega  }{\sum_p \int  C d \omega },
\label{eq:TTGgraballistic}
\end{align}
where the dimensionless temperature is determined by the frequency-dependent specific heat $C$ and group velocity $v_g$.

The analytical results are plotted in~\cref{TTGgarpheneBallistic}. An interesting phenomenon can be found, namely, at 10 K there is negative dimensionless temperature in the first trough, but this phenomenon disappears at room temperature.
The underlying physics is given below.
{\color{black}{The acoustic phonons contribute most heat conduction in graphene and the optical phonons contribute little to thermal transport due to high phonon scattering rates and small group velocities~\cite{pop_varshney_roy_2012,RevModPhys.90.041002}.
In low temperature, acoustic phonons with small wave vector have large group velocity and specific heat, see~\cref{garpheneparameters}(b, c), and the ZA mode contributes most thermal conduction~\cite{PhysRevBLindsay10} so that the thermal behaviors of only phonon ZA branch are the same as those of full phonon branches in graphene, as shown in~\cref{TTGgarpheneBallistic}.}}
{\color{black}{The ZA branch in small wave vector limit is a lower convex quadratic function~\cite{pop_varshney_roy_2012,RevModPhys.90.041002}, so that the results of Eq.~\eqref{eq:TTGgraballistic} could be negative.}}
{\color{black}{On the contrary, at room temperature, acoustic phonons with large wave vector have higher specific heat and the thermal contribution of ZA mode decreases~\cite{PhysRevBLindsay10} so that the thermal behaviors of only phonon ZA branch deviate from those of full phonon branches~\cite{pop_varshney_roy_2012,RevModPhys.90.041002}.}}

The numerical results obtained from the frequency-dependent BTE with different reference temperatures $T_0$ and grating period $L$ are shown in~\cref{TTGgarphene}.
When $L = 10~\mu$m, the grating period is much longer than the phonon mean free path at room temperature and the R-scattering dominates the heat conduction. Therefore, the temperature decreases exponentially with time and the heat wave disappears.
When $T_0$ decreases from the room temperature to $10$ K, phonons follow the diffusive, hydrodynamic and ballistic processes~\cite{cepellotti_phonon_2015,lee_hydrodynamic_2015,luo2019}.
The negative dimensionless temperature appears in low temperatures, as shown in~\cref{TTGgarphene}(a), which indicates that the heat wave could appear in both the hydrodynamic and ballistic regimes.
{\color{black}{When $L =1 ~\mu$m and $T_0 \leq 300$ K, the heat conduction is in the quasiballistic and ballistic regimes~\cite{cepellotti_phonon_2015,lee_hydrodynamic_2015,luo2019}.
From~\cref{TTGgarphene}(b), it can be found that the heat wave appears in this grating period regardless of temperature, which indicates that the heat wave could appear as long as the R scattering is not sufficient.}}

In summary, we find that the emergence of a heat wave in the ballistic regime depends on the frequency-dependent group velocity and specific heat.
In the suspended graphene, the heat wave appears in the hydrodynamic regime, but not necessary in the ballistic regime.

\subsection{Three-dimensional materials}

\begin{figure*}[htb]
	\centering
	\subfloat[R-scattering only] { \includegraphics[scale=0.28,clip=true]{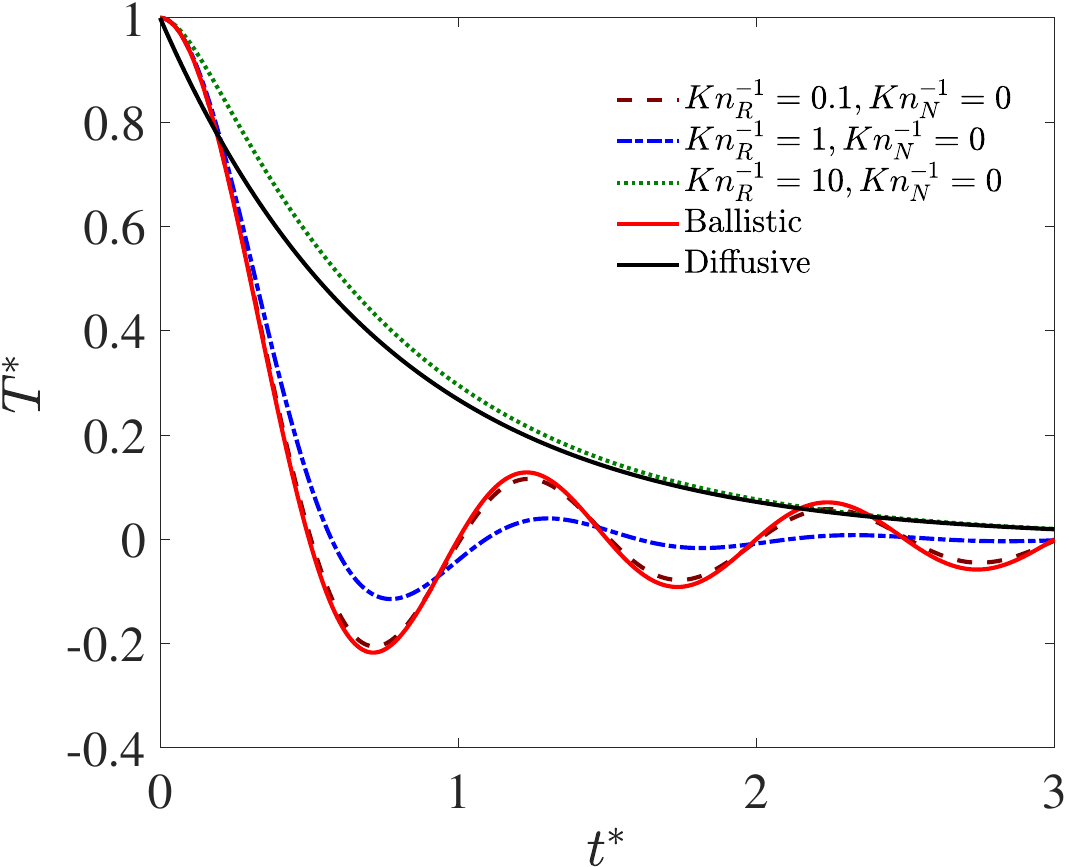} }
	\subfloat[N-scattering only] { \includegraphics[scale=0.28,clip=true]{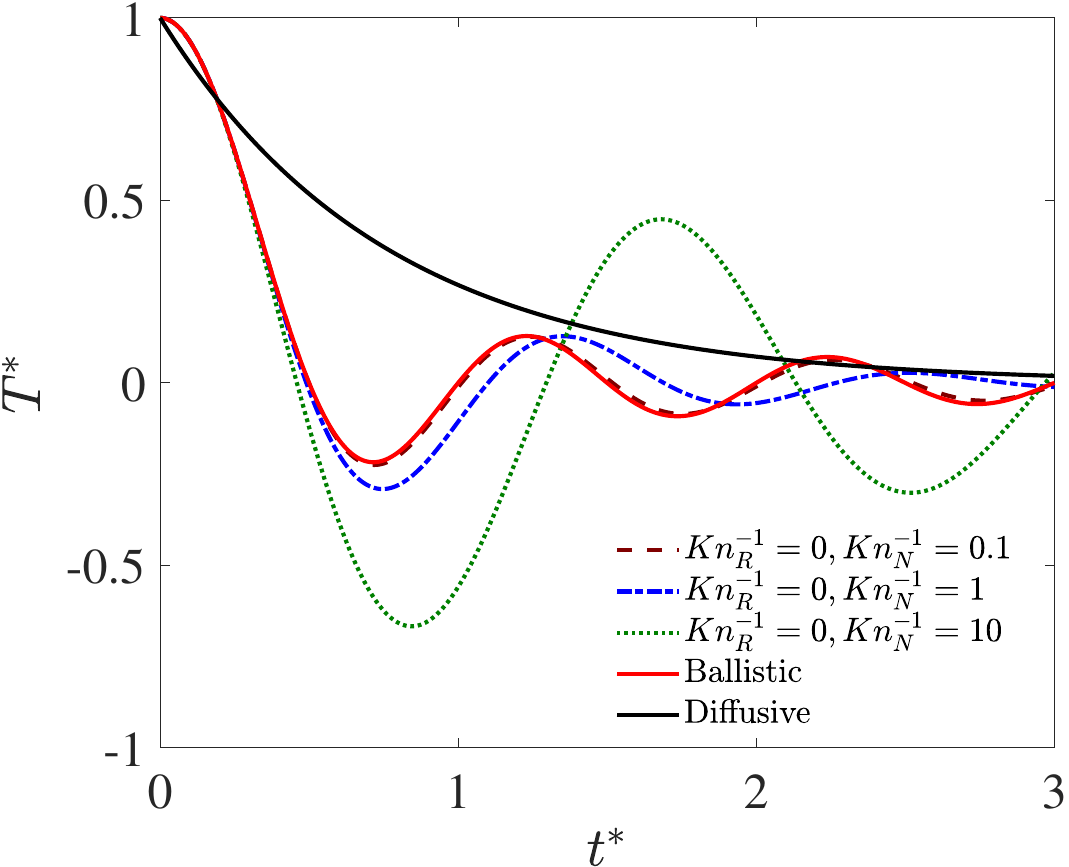} }
	\subfloat[Both N/R-scattering] { \includegraphics[scale=0.28,clip=true]{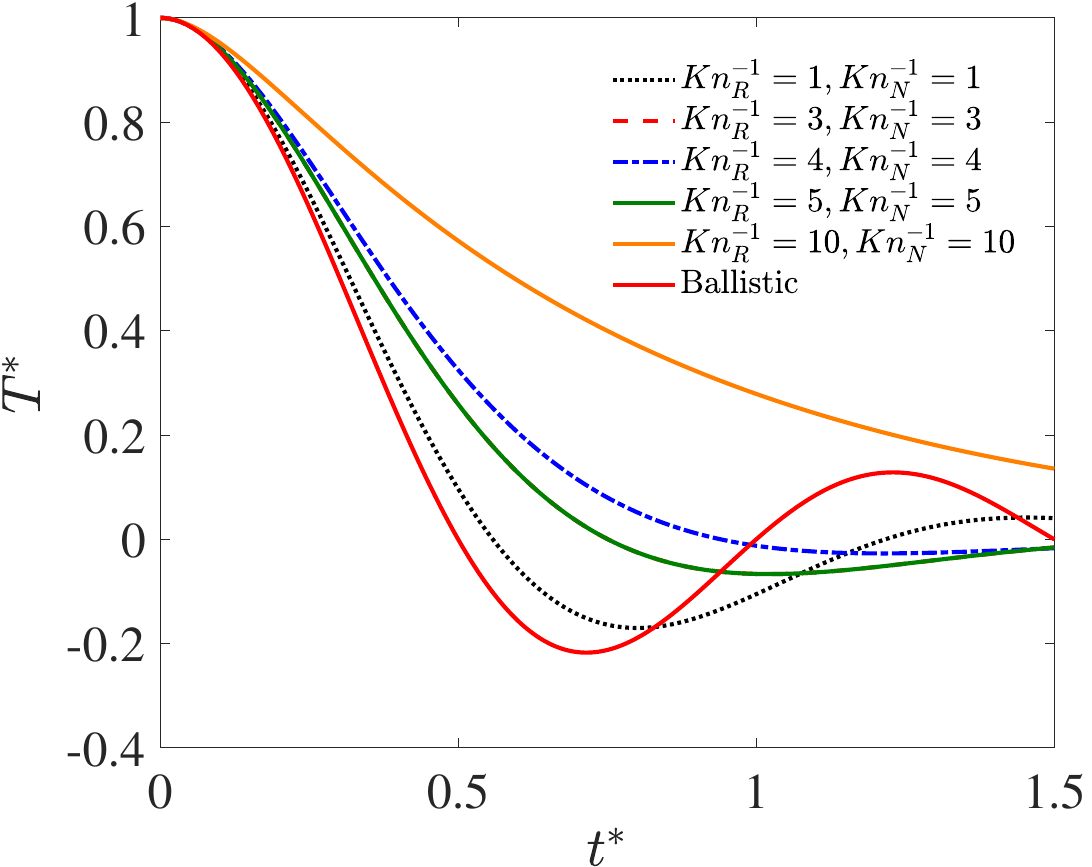} }
	\caption{The evolution of the dimensionless temperature~\eqref{dimensionless_temp} in the TTG simulations of three-dimensional materials based on frequency-independent BTE, where $t^*=v_g t/L$. Diffusive: Eq.~\eqref{eq:diffusive} with thermal conductivity $\kappa=Cv_g^2 \tau_R/3$. Ballistic: Eq.~\eqref{eq:TTG3Dballistic}. }
	\label{TTG3DgrayUN}
\end{figure*}
\begin{figure}[htb]
\centering
\subfloat[$L=100$ nm, $T_0=10$ K]{\includegraphics[scale=0.22,clip=true]{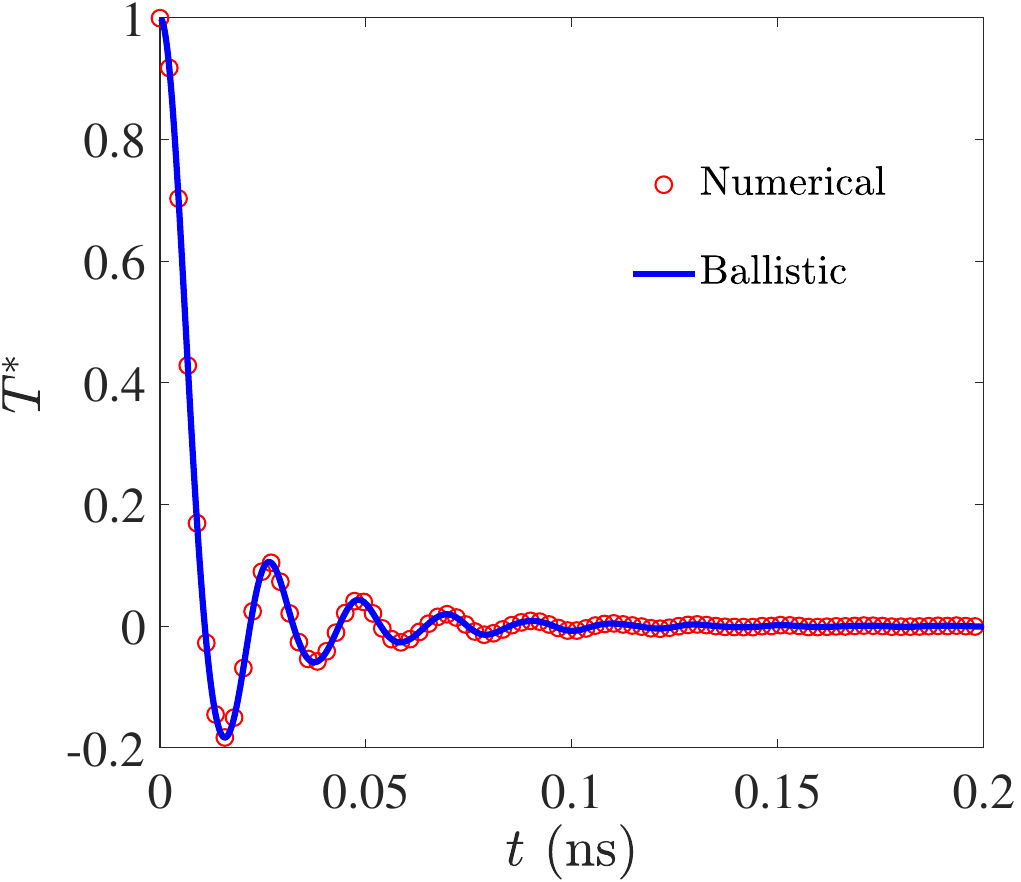} }
\subfloat[$L=100~\mu$m, $T_0=10$ K]{\includegraphics[scale=0.22,clip=true]{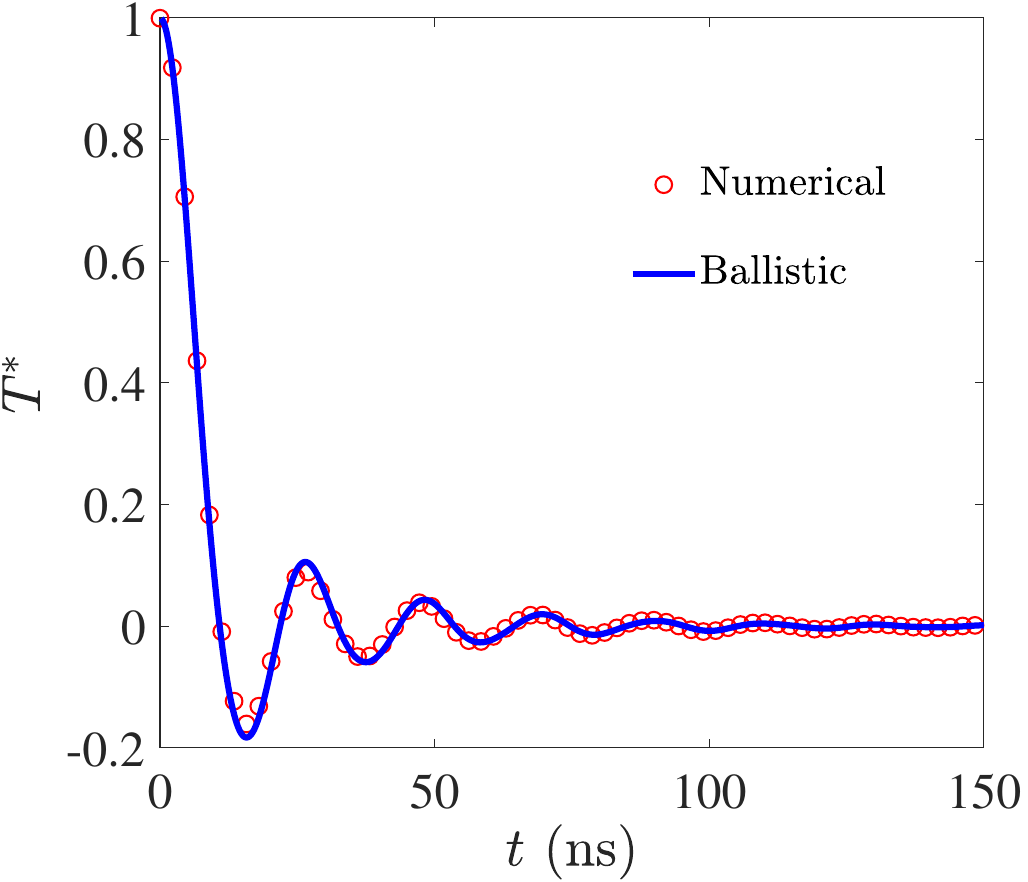}}  \\
\subfloat[$L=100$ nm, $T_0=30$ K]{\includegraphics[scale=0.22,clip=true]{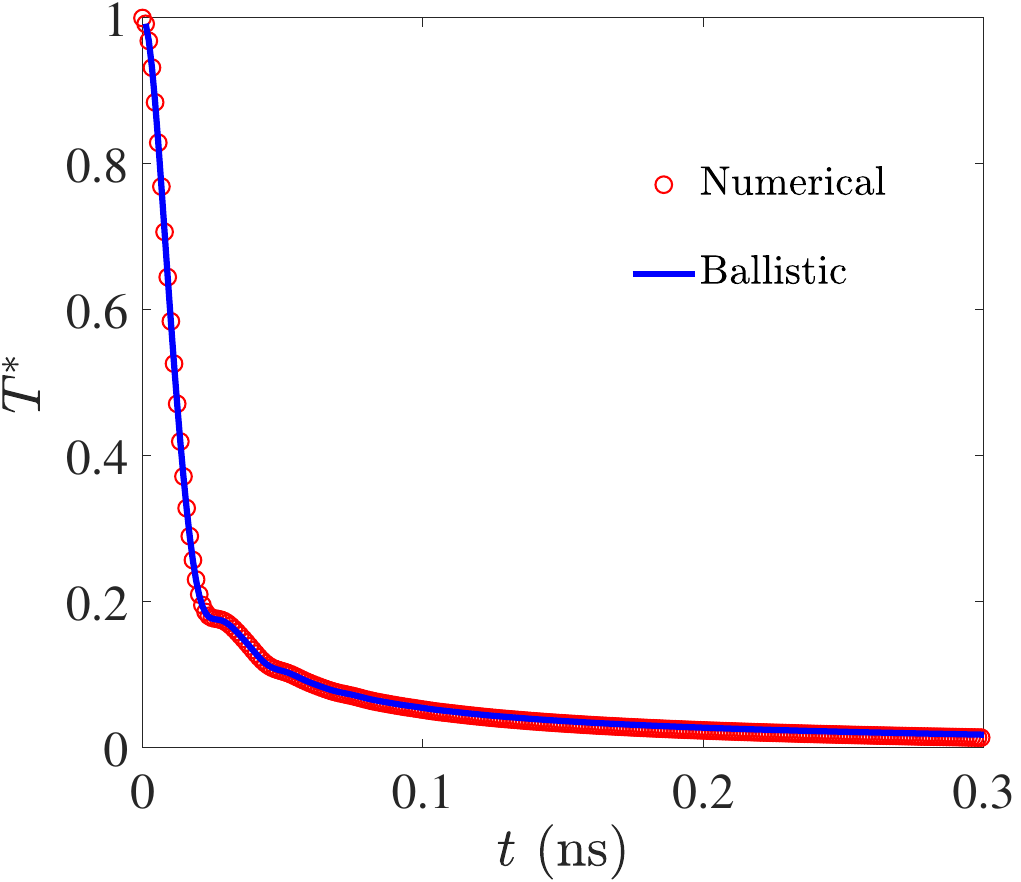} }
\subfloat[$L=100~\mu$m, $T_0=30$ K]{\includegraphics[scale=0.22,clip=true]{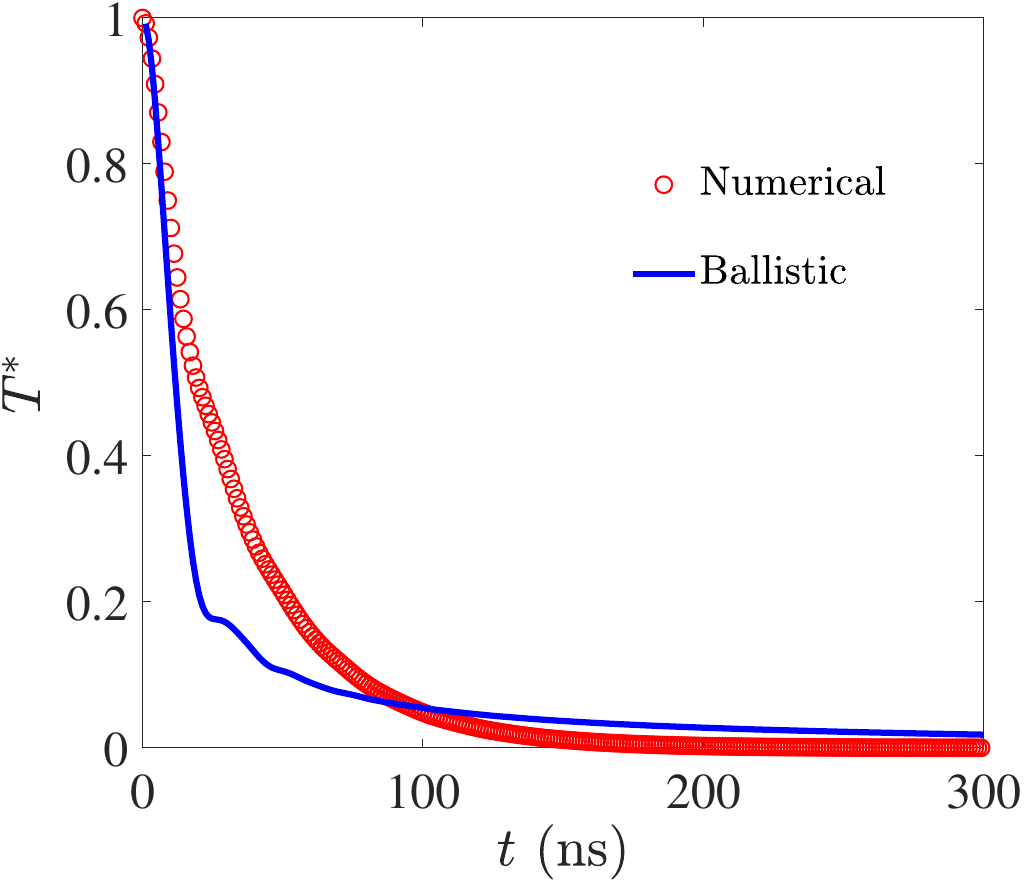}}  \\
\subfloat[$L=100$ nm, $T_0=300$ K]{\includegraphics[scale=0.22,clip=true]{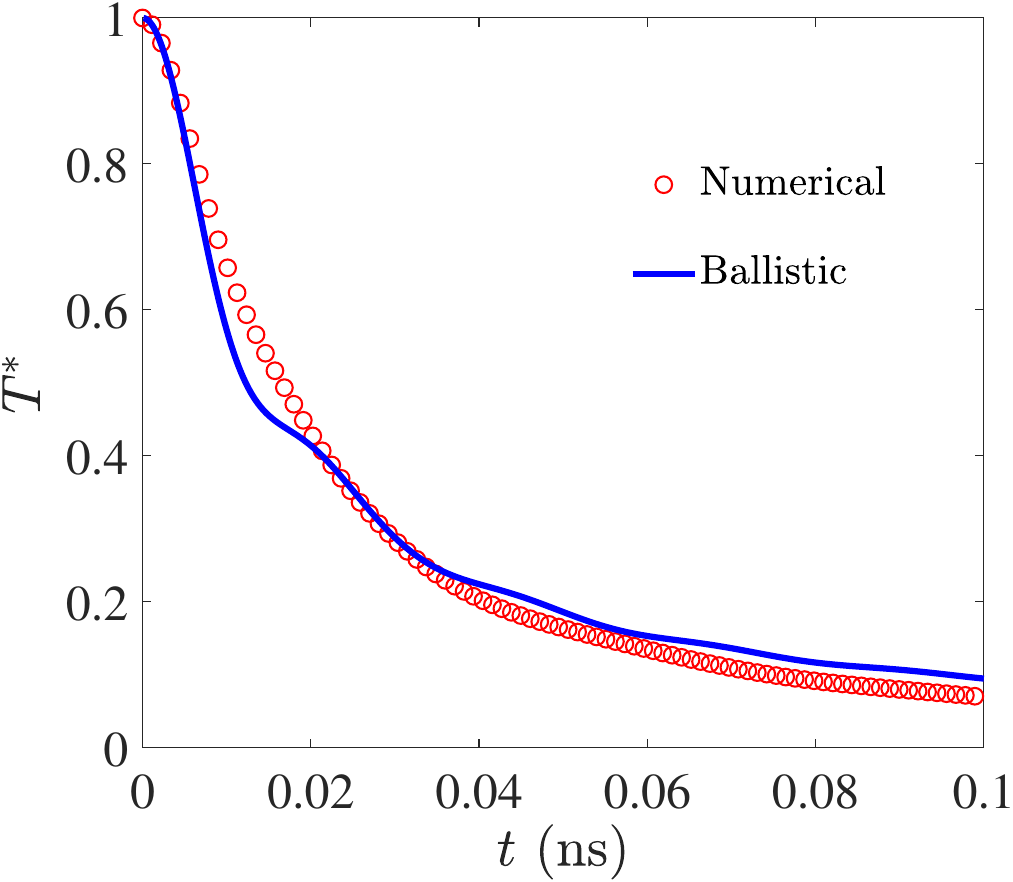}}
\subfloat[$L=100~\mu$m, $T_0=300$ K]{\includegraphics[scale=0.22,clip=true]{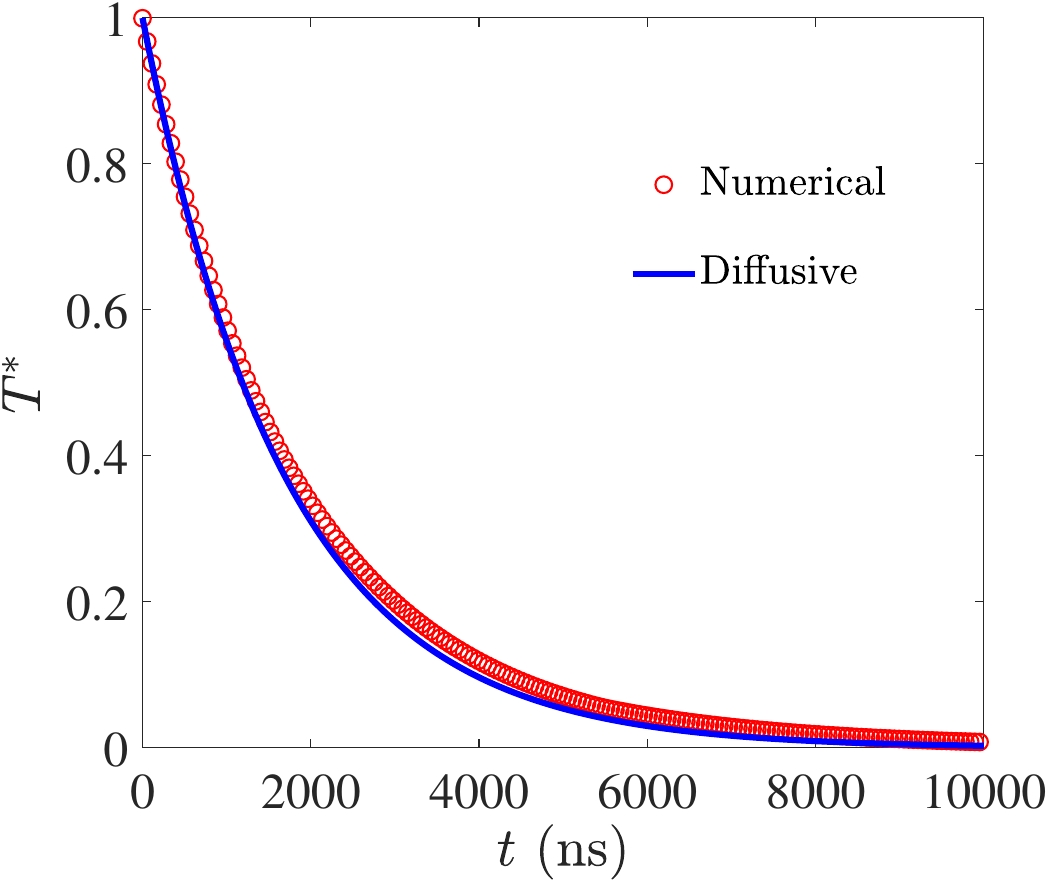}}
\caption{The evolution of the dimensionless temperature~\eqref{dimensionless_temp} in silicon with different reference temperatures $T_0$ and grating period $L$. Ballistic: Eq.~\eqref{eq:TTGSiballistic}. Diffusive: Eq.~\eqref{eq:diffusive} with thermal diffusivity $1.48$ cm$^2$s$^{-1}$. }
\label{TTGsilicon}
\end{figure}

To begin, the frequency-independent BTE is investigated. In addition to the analytical results obtained for the hydrodynamic and ballistic regimes, numerical solutions to the transition between regimes are shown in Fig.~\ref{TTG3DgrayUN}.
We found that the temperature response of three-dimensional materials is qualitatively similar to those in two-dimensional materials, i.e.,
the heat wave appears in the quasiballistic, ballistic and hydrodynamic regimes.

{\color{black}{Moving to the frequency-dependent BTE~\eqref{eq:dimensionlessBTE},}} the phonon dispersion and polarization of monocrystalline silicon are given in~Appendix~\ref{sec:dispersionsilicon}.
The thermal conductivity of bulk silicon is $145.9$ W/(m·K) at $300$ K~\cite{terris2009modeling}.
Distinct from graphene, there is no quadratic ZA branch~\cite{brockhouse1959lattice,pop2004analytic,kaviany_2008} and the distinction between N and R processes can be safely ignored in silicon across a wide range of temperatures~\cite{terris2009modeling}.

The numerical results for different system sizes and temperatures are shown in~\cref{TTGsilicon}.
{\color{black}{When $T_0=10$ K, as shown in~\cref{TTGsilicon}(a,b), the grating period is much smaller than the phonon mean free path and the numerical results agree well with the following analytical solutions in the ballistic regime:
\begin{align}
T^*(\bm{x}, t)  =\frac{   \sum_p \int  C \frac{ \sin( 2 \pi v_g t/L) }{2 \pi v_g t/L}      d\omega   }{   \sum_p \int  C d \omega   }.
\label{eq:TTGSiballistic}
\end{align}
A negative dimensionless temperature appears at $T_0 =10$ K, which indicates that a heat wave can emerge at extremely low temperatures.
Furthermore, at sufficiently low temperatures, the TA branch is linear, and the LA branch is considerably less important than TA branch due to its low density of state at low phonon frequency~\cite{kaviany_2008,ChenG05Oxford}.
In other words, heat conduction at extremely low temperatures in silicon can be approximated by the frequency-independent phonon BTE, as shown in~\cref{TTG3DgrayUN}.}}

{\color{black}{When $T_0 =30 $ K and $L=100$ nm, as shown in~\cref{TTGsilicon}(c), the negative dimensionless temperature disappears although the phonon transport is still in the ballistic regime.
Note that although there are slight temperature oscillations, but no negative dimensionless temperature, for the ballistic curves, it is simply the oscillation of the amplitude, not the phase lag, which indicates that there is no wave-like propagation of heat along the $x$ direction.
Therefore, heat waves are not a necessary signature for the ballistic regime in three-dimensional materials.}}
When the grating period is comparable to the phonon mean free path, quasiballistic phonon transport becomes the dominant regime of transient heat conduction, as shown in \cref{TTGsilicon}(d,e), the behaviors of dimensionless temperature deviate from the analytical solutions in the ballistic limit.
When the grating period is $L=100~\mu$m, phonons undergo diffusive transport and correspondingly follows Fourier's law of thermal conduction at room temperature, see~\cref{TTGsilicon}(f).

\section{CONCLUSION}
\label{sec:conclusion}

{\color{black}{The temperature response in the quasi-one dimensional transient thermal grating heating geometry in the ballistic, quasi-ballistic and hydrodynamic phonon transport regimes have been investigated based on analytical and numerical solutions to the Callaway approximation to the phonon BTE.
Our results uncover the interplay between a material's phonon dispersion relation, the relative intensity of N/R scattering  and the background temperature contributing to the emergence of heat waves.
For the frequency-independent BTE, heat waves will appear so long as R scattering is not sufficient relative to the grating period (quasiballistic and ballistic) or to the N scattering rate (hydrodynamic).
However, for the frequency-dependent BTE, heat waves manifest in the phonon hydrodynamic regime but do not necessarily appear in the ballistic regime.
Ultimately, the phonon dispersion relation and temperature determine whether a heat wave will manifest in the ballistic regime. As shown, in suspended graphene and silicon, heat waves are predicted to appear at extremely low temperatures but subsequently disappear at room temperature and above.}}

\section*{ACKNOWLEDGMENTS}

This study is supported by National Natural Science Foundation of China (12147122) and the China Postdoctoral Science Foundation (2021M701565).
The authors acknowledge Dr. Andrea Cepellotti, Albert Beardo and Manyu Shang for useful communications on driftless second sound.
\textbf{
The authors are grateful to the anonymous referees for their valuable suggestions.}

\appendix

\section{Phonon dispersion and scattering in silicon}
\label{sec:dispersionsilicon}

The phonon dispersion relations of the silicon in the [100] direction are chosen to represent the other directions~\cite{brockhouse1959lattice,kaviany_2008}.
Only the acoustic phonon branches (LA, TA) are considered because the optical branches contribute little to the thermal conduction.
The dispersion relations of the acoustic phonon branches can be approximated by quadratic polynomial dispersions~\cite{pop2004analytic}:
\begin{equation}
\omega=c_{1}k+c_{2}k^2,
\label{eq:curves}
\end{equation}
where the wave vector $k \in[0,k_{\text{max}}]$, $k_{\text{max}}=2\pi /A $ is the maximum wave vector in the first Brillouin zone, $A$ is the lattice constant.
For silicon, $A=0.543$ nm.
For LA, $c_1= 9.01\times 10^5$ cm/s, $c_2= -2.0 \times 10^{-3}$ cm$^2$/s.
For TA, $c_1= 5.23\times 10^5$ cm/s, $c_2= -2.26 \times 10^{-3}$ cm$^2$/s.
The phonon group velocity is
\begin{equation}
v_g = \frac{\partial \omega }{ \partial  k  } = c_1 +2 c_2 k.
\end{equation}

The Terris' rule~\cite{terris2009modeling} is used to calculate the relaxation time: $\tau_R^{-1}=\tau_{{\text{impurity}}}^{-1}+\tau_{{\text{U}}}^{-1}+\tau_{{\text{N}}}^{-1}=\tau_{{\text{impurity}}}^{-1}+\tau_{{\text{NU}}}^{-1}$, where $\tau_{{\text{impurity}}}^{-1}  =  A_{i}\omega^{4}$, $A_{i}=1.498\times10^{-45}~{\text{s}^{\text{3}}}$.
For LA, $\tau_{{\text{NU}}}^{-1}=B_{L}\omega^{2}T^{3}$, where $B_{L}=1.180\times 10^{-24}~{\text{K}^{\text{-3}}}$.
For TA, when $0 \leq k < k_{max}/2$, $\tau_{{\text{NU}}}^{-1}=B_T\omega T^4$ and when $k_{max}/2 \leq k \leq k_{max}$, $\tau_{{\text{NU}}}^{-1}=B_U\omega^{2}/{\sinh(\hbar\omega/k_{B}T)}$, where $B_T=8.708\times 10^{-13}~{\text{K}^{\text{-3}}}$ and $B_{U}=2.890\times10^{-18}~{\text{s}}$.
The normal processes are ignored in silicon, that is $\tau_N^{-1}=0$.

\section{Numerical method for BTE}
\label{sec:solver}

$Solvers$---The discrete unified gas kinetic scheme invented by Guo~\cite{guo_progress_DUGKS,GuoZl16DUGKS} is used to solve the frequency-independent phonon BTE numerically.
Detailed introductions and numerical validations of this scheme can refer to previous studies~\cite{luo2019,zhang_transient_2021}.
The frequency-dependent phonon BTE~\eqref{eq:BTE} is solved numerically by the implicit scheme:
\begin{equation}\label{eq:dualimplicit}
\begin{aligned}
& \frac{ e^{n+1}  -e^{n} }{ \Delta t} + v_g \bm{s} \cdot \nabla e^{n+1}  \\
=  &  \frac{ e^{eq}_{R}(T_R^{n+1} ) -e^{n+1} }{\tau_{R}} + \frac{e^{eq}_{N}(T_N^{n+1} )-e^{n+1}}{\tau_{N}},
\end{aligned}
\end{equation}
where $n$ represents the time step, $\Delta t$ is the physical time step.
$T_R$ and $T_N$ are the pseudo-temperatures, which are introduced to ensure the conservation properties of phonon N- and R-scattering.
In order to solve Eq.~\eqref{eq:dualimplicit}, an iterative scheme is used:
\begin{equation}
\begin{aligned}
& \frac{ e^{n,m}  -e^{n} }{ \Delta t} + v_g \bm{s} \cdot \nabla e^{n,m}  \\
= &  \frac{ e^{eq}_{R}(T_R^{n,m} ) -e^{n,m } }{\tau_{R}} + \frac{e^{eq}_{N}(T_N^{n,m} )-e^{n,m}}{\tau_{N}},
\label{eq:iterations}
\end{aligned}
\end{equation}
where $m \geq 0 $ is the index of inner iteration. When $m=0$, $e^{n,m} =e^n$, $T_N^{n,m} =T_N^n$, $T_R^{n,m} =T_R^n$.
When the inner iterations converge ($m \rightarrow  \infty $), we set $e^{n,m} =e^{n+1}$, $T_N^{n,m} =T_N^{n+1}$, $T_R^{n,m} =T_R^{n+1}$.

$Discretizations$---The spatial space is discretized with $N_{cell}=400$ uniform cells.
For all cases, the first-order upwind scheme is used to deal with the spatial gradient of the distribution function in the ballistic regime, and the van Leer limiter is used in the hydrodynamic and diffusive regimes.
The time step is
\begin{align}
\Delta t= \text{CFL} \times \frac{\Delta x}{v_{\text{max}} },
\label{eq:CFLnumber}
\end{align}
where $\Delta x$ is the minimum discretized cell size, $\text{CFL}$ is the Courant–Friedrichs–Lewy number and $v_{\text{max}} $ is the maximum group velocity.
In this simulations, $\text{CFL}=0.04-0.4$.

For two-dimensional materials, we set $\bm{s}=\left( \cos \theta, \sin \theta  \right)$, where $\theta \in [0, 2 \pi]$ is the polar angle.
Due to symmetry, the $\theta \in [0,\pi]$ is discretized with the $N_{\theta}$-point Gauss-Legendre quadrature.
The total number of the discretized directions is $2N_{\theta}$.
In this study, we set $N_{\theta}=50$.
In graphene, the discretized phonon dispersion and polarization are shown in~\cref{garpheneparameters} and total $300$ discretized frequency bands are considered.
Besides, the mid-point rule is used for the numerical integration of the frequency space.
For three-dimensional materials, $\bm{s}=\left( \cos \theta, \sin \theta \cos \varphi, \sin \theta  \sin \varphi  \right)$, where $\theta \in [0,\pi]$, $\varphi \in [0,2\pi]$ is the azimuthal angle.
The $\cos \theta \in [-1,1]$ is discretized with the $N_{\theta}$-point Gauss-Legendre quadrature, while the azimuthal angular space $\varphi \in [0,\pi]$ (due to symmetry) is discretized with the $\frac{N_{\varphi}}{2}$-point Gauss-Legendre quadrature.
In this study, we set $N_{\theta} \times N_{\varphi} =100 \times 8$.
In silicon, the wave vector is discretized equally and the mid-point rule is used for the numerical integration of the frequency space.
Total $40$ discretized frequency bands are considered, which is enough to ensure the convergence of bulk thermal conductivity.
The present discretizations of the wave vector space are enough for all phonon transport regimes;
one example is given in~\cref{TTGBD}.

\bibliography{phonon}

\end{document}